\documentclass[doublecol]{epl2}
\usepackage{graphicx}
\usepackage{amssymb}
\usepackage{amsmath}
\usepackage{latexsym}
\usepackage{color}

\title{Magnetic spin excitations in Mn doped GaAs: A model study}
\author{Akash Chakraborty\inst{1} \and Richard Bouzerar\inst{1} \and Georges Bouzerar\inst{1,2}}
\shortauthor{Akash Chakraborty \etal}

\institute{
  \inst{1} Institut N\'eel, CNRS, D\'epartement MCBT, 25 avenue des Martyrs, B.P. 166, 38042 Grenoble Cedex 09, France \\
  \inst{2} School of Engineering and Science, Jacobs University Bremen, Campus Ring 1, D-28759 Bremen, Germany
}


\abstract{
We provide a quantitative theoretical model study of the dynamical magnetic properties of optimally annealed Ga$_{1-x}$Mn$_x$As. This model has already been shown to reproduce
accurately the Curie temperatures for Ga$_{1-x}$Mn$_x$As. Here we show that
the calculated spin stiffness are in excellent agreement with those which were obtained from ab-initio based studies.
In addition, an overall good agreement is also found with available experimental data.
We have also evaluated the magnon density of states and the typical density of states from which the ``mobility edge'', separating the extended from localized magnon states, was
determined. The power of the model lies in its ability to be generalized for a broad class of diluted magnetic semiconductor materials, thus it bridges the gap between
first principle calculations and model based studies.
}

\begin{document}

\maketitle

The prospect of manipulating the electronic spin for spintronics applications has generated
a tremendous interest in the so called diluted magnetic semiconductors (DMS) over the past few years\cite{satormp,jungwirth}. Among the III-V semiconductors, Mn doped GaAs is the most widely
studied. The understanding of its fundamental physical properties involves large theoretical speculations\cite{dietl,inoue,schliemann}, like
there is still a controversy regarding the existence of a preformed
impurity band (IB) in GaMnAs. Most of the model studies are based on a perturbative treatment of Mn in GaAs (the valence band (VB) picture)\cite{jungwirth,dietl}.
This is in contradiction to the ab-initio calculations\cite{bergqvist2,sandratskii}, which shows that Mn strongly affects the nature of the states close to
the Fermi level ($E_F$) leading at low impurity concentration to a preformed impurity band. This supports the IB picture and rules out the VB picture. Note also that (Ga,Mn)As is close to the metal insulator transition. Indeed after
annealing as grown samples an insulator-metal transition was also often observed experimentally\cite{matsukura1,hayashi}. This important feature cannot be captured in the framework of the perturbative VB theory.

 Infrared and optical spectroscopy measurements have also shown that $E_F$ resides indeed in a Mn induced IB in GaMnAs\cite{burch}. In the most frequently studied model,
 based on a six or eight band Kohn-Luttinger Hamiltonian\cite{jungwirth,dietl}, the pd-coupling is treated perturbatively and the dilution effects are neglected (the VB picture),
which as aforementioned are inconsistent with the ab-initio based studies. It has been demonstrated that the perturbative treatment is
 inappropriate for DMS and that both thermal fluctuactions and disorder effects play a crucial role. The importance of treating disorder effects in a reliable manner was often
underlined \cite{bhatt,richard,richard1}. The first principle based calculations were able to reproduce the Curie temperatures accurately\cite{georgesapl,bergqvist1,georges2} but
 they are essentially material specific. Hence the ideal tool to identify and analyze the effect of the relevant physical parameters is a minimal model approach based on a non-perturbative treatment of the substitution effects.

In this communication we use the one band V-J model\cite{richard1}, treated non-perturbatively, to study the dynamical magnetic properties of the III-V DMS GaMnAs. It will be shown
that the V-J model provides the missing link between ab-initio and model studies.
To be more specific we adopt a two-step approach to calculate the magnetic properties of GaMnAs. First the couplings between the magnetic impurities are calculated within the
one band V-J model without the use of any effective medium theory. In the second step
 the effective  dilute Heisenberg Hamiltonian is treated within the self consistent local RPA (SC-LRPA) theory to calculate the magnetic properties as well as the T$_C$s.
Note that the accuracy and reliability of SC-LRPA to treat dilution/disorder and thermal fluctuations has been proved time and again\cite{georges2,akash}.
The non-perturbative treatment of the V-J model was shown to evaluate qualitatively the essential magnetic properties of a whole class of III-V materials\cite{richard1}.
Recently the model was also shown to successfully reproduce and explain quantitatively the measured optical conductivity in
Ga$_{1-x}$Mn$_x$As\cite{georges1}. In the present work, we calculate the T$_C$s, the magnon density of states (DOS) (average and typical) along with the ``mobility edge'', the magnon spectral
function and the spin stiffness as a function of the impurity concentration in optimally annealed Ga$_{1-x}$Mn$_x$As. The term optimally annealed implies that the concentration of
compensating defects (As anti-sites or Mn interstitials) is negligible. Note that here we focus on homogeneously diluted systems with no correlations in
 disorder (absence of inhomogeneities).

\vspace{0.5cm}
\begin{figure}[htbp]{
\includegraphics[width=8cm,angle=0]{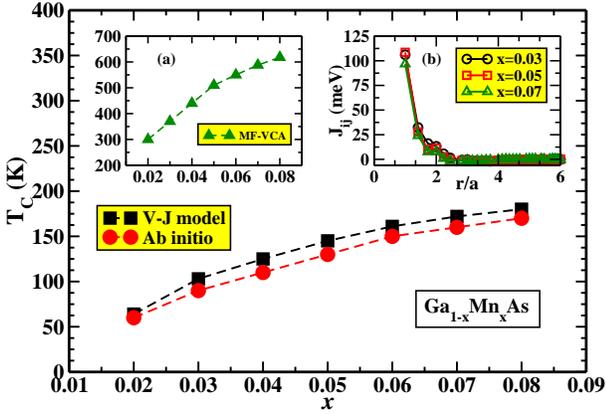}}
\caption{(Color online) Curie temperature (in K) as a function of the Mn concentration ($x$). The squares represent our model and the circles correspond to
 ab-initio calculations. Inset (a) shows the MF-VCA T$_C$ (in K) as a function of $x$. Inset (b) shows the exchange couplings (in meV) as a function of
distance for three different values of $x$.}
\label{fig.1}
\end{figure}
The one band V-J Hamiltonian describing the interaction between the carriers (holes or electrons) and the localized impurity spins is given by,
\begin{equation}
\label{eq.1}
H=-\sum_{ij,\sigma} t_{ij}c_{i\sigma}^{\dagger}c_{j\sigma}+
\sum_{i} J_{i}{\bf S}_{i}\cdot {\bf s}_{i}+
\sum_{i\sigma} V_{i}c_{i\sigma}^{\dagger}c_{i\sigma}
\end{equation}

where the hopping term t$_{ij}$=t for $i$ and $j$ nearest neighbors, otherwise zero. c$_{i\sigma}^{\dagger}$ (c$_{i\sigma}$) is the creation (annihilation) operator
of a hole of spin $\sigma$ at site $i$. J$_i$ is the p-d coupling (J$_{pd}$) between localized Mn spin ${\bf S}_i$ ($|{\bf S}_{i}|$=5/2) and a spin carrier
 ${\bf s}_i$ (p-band). The on-site potential V$_{i}$ results from the substitution of Ga$^{3+}$ by Mn$^{2+}$. Note that V$_{i}$ is directly related to the position
of the bound state with respect to the VB. J$_i$=p$_i$J$_{pd}$ and V$_i$=p$_i$V, where p$_i$=1 if the site is occupied by an impurity, otherwise zero. Here $x$ and
 $p$ represent the Mn concentration and hole density respectively. The next task is to fix the model parameters. Since our calculations are performed on a simple cubic lattice 
(1 atom per unit cell) the value of the lattice parameter used here is $a$=$\frac{a_0}{4^{1/3}}$=3.55\AA, where $a_0$=5.65\AA  is the lattice
constant for zinc-blende GaAs (4 Ga per unit cell). The hole effective mass for density of states in GaAs is known to be $\sim$ 0.5$m_e$\cite{book}, from which the value of t is fixed to 
0.7 eV. Note that a variation of about $\pm$10\% of t does not affect our results considerably.  J$_{pd}$S is set to the value of 3 eV since J$_{pd}\approx$1.2 eV\cite{okabayashi}
 is already known for GaMnAs. The last parameter, the on-site potential V is the crucial one. It is set to 1.8t in order to reproduce
the bound hybridized pd-states energy E$_b\approx$110 meV in GaAs host\cite{chapman} with respect to the top of the VB. Now in the realistic material each Mn$^{2+}$ provides one 
hole and n$_l$=3 degenerate p-d states near the top of the VB (here $p$=$x$ for the well annealed case). Thus for reasons of consistency our model calculations for optimally annealed
samples are performed for the hole density $\bar{p}$=$p$/n$_l$. Then in both cases the IB is one-third filled (for more details see Ref.\cite{georges1,richard2}).
Let us now proceed with the calculation of the exchange couplings from our one band model. For a given impurity concentration
`$x$' and disorder configuration `$c$', the Hamiltonian (1) is diagonalized exactly in both spin sectors. We have performed the calculations on simple cubic systems
 with sizes varying from $L$=16 to $L$=24 and the average over disorder is done for a few hundred configurations. A careful study of the finite size effects on the magnetic
couplings has been made. The diagonalization provides us with the eigenvalues and eigenvectors denoted by \{$\omega^c_{\sigma,\alpha},|\Psi^c_{\sigma,\alpha}\rangle$\},
where $\sigma$=$\uparrow,\downarrow$ and $\alpha$=1,2,...,N (N=$L^3$ is the total number of sites), which are used to evaluate the couplings. The magnetic coupling
between two localized spins is given by the generalized susceptibility\cite{katsnelson}
\begin{equation}
\label{eq.2}
\bar{J}_{ij}(x,\bar{p})=-\frac{1}{4\pi{S^2}}\Im\int_{-\infty}^{E_F}Tr({\Sigma_i}{G_{ij}^\uparrow(\omega)}{\Sigma_j}{G_{ji}^\downarrow(\omega)})d\omega
\end{equation}
where the Green's functions are defined as G$_{ij}^\sigma(\omega)$=
$\langle i{\sigma}|\frac{1}{\omega-\hat{H}+i\epsilon}|j{\sigma}\rangle$. It is found that the couplings calculated for well annealed Ga$_{1-x}$Mn$_x$As are rather
short range and essentially ferromagnetic. These are similar in nature to those obtained from first principle studies\cite{richard1,kudrnovsky}. Now the couplings that are
used to calculate the T$_C$s and other magnetic properties are defined as $J_{ij}(x,p)$=$n_l\bar{J}_{ij}(x,\bar{p}=p/n_l)$, since we stress again that
 in our one band model each Mn$^{2+}$ provides a single state unlike the 3 p-d states in the realistic case. The exchange couplings for three different Mn concentrations
are shown as a function of distance in inset (b) of fig.~\ref{fig.1}. Note that the same argument was used to calculate the transport
 properties, namely the optical conductivity, in Ga$_{1-x}$Mn$_x$As and excellent agreement with experimental results was obtained \cite{georges1}. As we have found that
the spin stiffness is sensitive to the size of the lattice on which the couplings are calculated, here the calculations have been performed systematically on considerably
larger systems compared to previous studies\cite{richard2} leading to an improvement in the couplings at large distances. It should be noted that
 calculating the exchange couplings and the corresponding magnetic quantities simultaneously for a given configuration and then averaging these obtained quantities over 
several configurations leads to the same results as the ones shown here.

\vspace{0.5cm}
\begin{figure}[htbp]{
\includegraphics[width=8.0cm,angle=0]{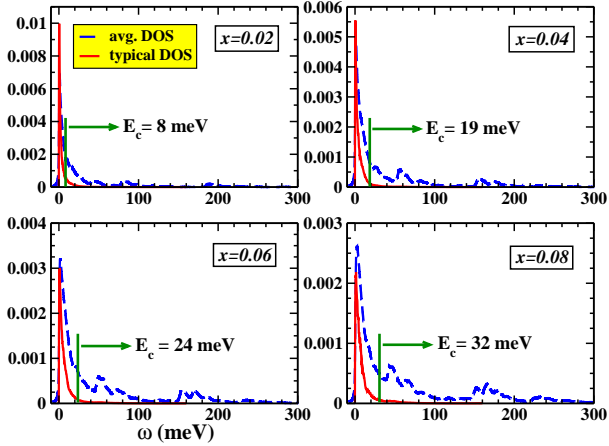}}
\caption{(Color online) Magnon density of states (along the $y$ axis) as a function of the energy $\omega$ (in meV) (along the $x$ axis) for different Mn concentration $x$.
The solid (dashed) line represents the typical (average) density of states.}
\label{fig.2}
\end{figure}

In fig.~\ref{fig.1} we show the Curie temperature (in Kelvin) as a function of the Mn concentration for well annealed samples. To evaluate the T$_C$,
the effective dilute Heisenberg Hamiltonian $H_{Heis}$=$-\sum_{i,j} p_{i}p_{j}J_{i,j}(x,p){\bf S}_{i}\cdot {\bf S}_{j}$, is treated within the SC-LRPA theory. After
performing the Tyablikov decoupling and diagonalizing the above Hamiltonian we obtain the retarded Green's functions given by
\begin{equation}
\label{eq.3}
\mathcal{G}_{ij}^c(\omega)=\sum_\alpha \frac{2\langle{S_j^z}\rangle}{\omega-\omega_\alpha^c+i\epsilon}
\langle i|\Phi_\alpha^c\rangle \langle\Phi_\alpha^c|j\rangle
\end{equation}
where $\Phi_\alpha^c$ and $\omega_\alpha^c$ are now the eigenvalues and eigenvectors respectively of $H_{Heis}$ for a given configuration `c', and
$\langle{S_j^z}\rangle$ is the local magnetization (for details see Ref.\cite {georges2}). The diagonalization of the above Hamiltonian was performed on very
 large systems (typically $L$=50) together with a systematic average over a few hundred configurations of disorder. We have found that the T$_C$s shown here are
comparable to those shown in Ref.\cite{richard2}. As is clearly seen from the  figure, our model calculations are in very good agreement with the T$_C$s obtained from first principle
studies\cite{bergqvist1,georges2,sato}, which in turn reproduced accurately the experimental data\cite{edmonds,matsukura,chiba}. We again stress the fact that there are no
adjustable parameters in our model calculations.
 It is worth noting that this model can also be used to evaluate the T$_C$s and the magnetic and transport properties for a whole class of DMS
materials, which makes it all the more powerful\cite{richard2}. Several attempts have been made to calculate T$_C$s using mean field theories\cite{zhao}
but they have always resulted in overestimations. As shown the mean field Virtual Crystal Approximation (MF-VCA) T$_C$ (inset (a) of fig.~\ref{fig.1}) already
leads to room temperature for only 2$\%$ of Mn. Once again it shows the clear overestimation in the T$_C$ values and hence the crucial importance of thermal fluctuations and
 dilution effects.

The average and typical magnon DOS as a function of the energy for different concentrations of Mn are plotted in fig.~\ref{fig.2}. The size of the simple cubic system in this
 case is $L$=44. The average magnon DOS is given by $\rho_{avg}(\omega)$=$\langle({1}/{N_{imp}})\Sigma_i\rho_i(\omega)\rangle_c$, where $\rho_i(\omega)$=$-({1}/{2\pi\langle S_i^z \rangle})$Im$\mathcal{G}_{ii}(\omega)$
is the local magnon DOS, $N_{imp}$ is the total number of magnetic impurities and $\langle...\rangle_c$
denotes the average over disorder configurations (for details see Ref.\cite{akash}). As seen from the figure, the shape of $\rho_{avg}(\omega)$
exhibits a multipeak structure and changes significantly with increasing $x$. At sufficiently high dilution, close to the percolation, there is a drastic
increase in the low energy DOS. This is due to formation of isolated impurity clusters at low concentrations which have their own zero-energy eigenmodes and these
 contribute to the average magnon DOS. However $\rho_{avg}(\omega)$ is not able to predict the nature of the magnon modes, whether they are extended or localized. For this
we calculate the typical magnon DOS\cite{nikolic} which is given by $\rho_{typ}(\omega)$=exp(${\langle}$ln ${\rho_i(\omega)\rangle}_c$). The typical magnon DOS is a local quantity and
 unlike $\rho_{avg}(\omega)$ it provides direct access to the ``mobility edge''
separating the localized modes from the extended ones. The solid (red) line in the figure corresponds to $\rho_{typ}(\omega)$ and apparently there is no significant change
 in the overall shape with variation in dilution. We have evaluated from $\rho_{typ}(\omega)$ the energy E$_c$, which separates the localized from the extended states. The
values are indicated in the figure. As we increase the Mn concentration E$_c$ varies from 8 meV for $x$=0.02 to 32 meV for $x$=0.08. This shows that even for
relatively larger concentrations most of the magnon excitations consists of localized states (fractons)\cite{fractons}.
\begin{figure}[htbp]{
\includegraphics[width=8cm,angle=0]{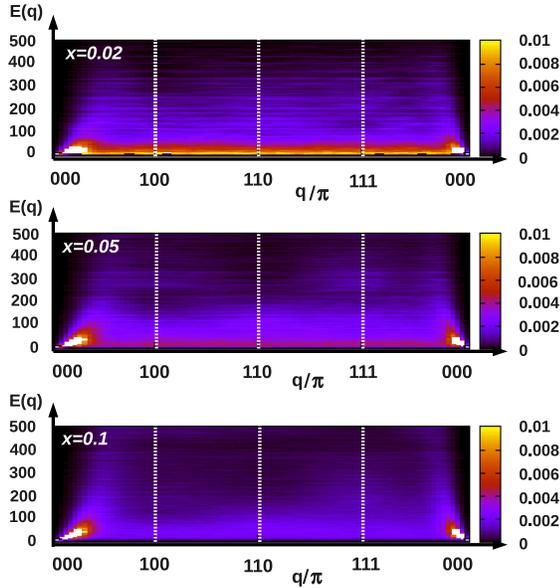}}
\caption{(Color online) Spectral function $A({\bf q},\omega)$ in the $({\bf q},\omega)$ plane for different Mn concentration ($x$). The energy axis ($y$ axis) is in meV.
(The size of the simple cubic lattice is $L$=44).}
\label{fig.3}
\end{figure}

In the next figure,fig.~\ref{fig.3} , we show the magnon spectral function $A({\bf q},\omega)$=$-({1}/{\pi\langle\langle{S^z}\rangle\rangle})$Im$\mathcal{G}({\bf q},\omega)$,
 in the $({\bf q},\omega)$ plane over the entire Brillouin zone corresponding to three different concentrations of Mn. The spectral function is directly accessible by
inelastic neutron scattering experiments and provides direct insight into the magnetic excitation spectrum (details can be found in Ref.\cite{akash}). The
 average was done over a few hundred configurations of disorder but we found that increasing the number of configurations  beyond 50 left the results almost unaffected. Now
 similar results were obtained for the excitation spectrum of Ga$_{1-x}$Mn$_x$As \cite{georges3} but it is to be noted the couplings in that case were calculated from
 first principle Tight Binding Linear Muffin Tin Orbital approach. We observe that well defined excitations exist only in a restricted region of the Brillouin zone
essentially around the $\Gamma$ point [$\bf {q}$=(000)]. The spectrum shows a significant broadening as we move away from the $\Gamma$ point. This is in agreement with the
 typical DOS calculations shown in the previous figure. Thus we can see that our results are in very good agreement with those obtained from first principle studies cited above.

\begin{figure}[htbp]{
\includegraphics[width=8cm,angle=0]{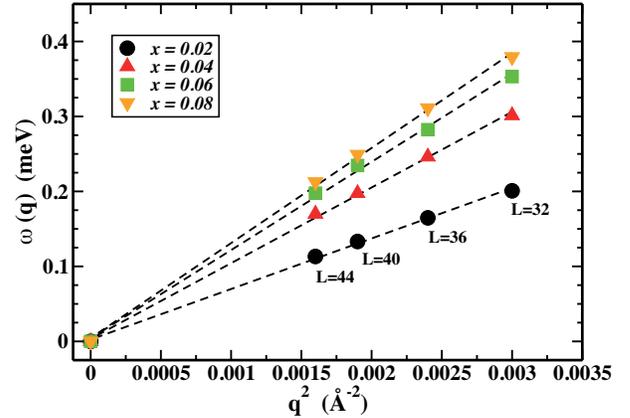}}
\caption{(Color online) Magnon energy $\omega(\rm{\bf q})$ (in meV) as a function of q$^2$ (in {\AA}$^{-2}$) for different concentration of Mn ($x$).
(L is the size of the simple cubic lattice).}
\label{fig.4}
\end{figure}

\begin{figure}[htbp]{
\includegraphics[width=8cm,angle=0]{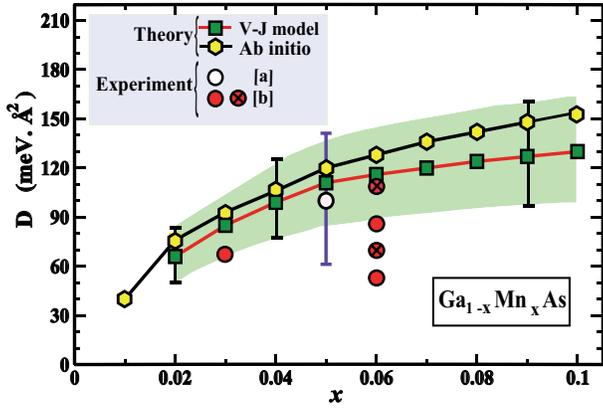}}
\caption{(Color online) Spin stiffness D (in meV.{\AA}$^{2}$) as a function of the Mn concentration ($x$). Squares correspond to our model, hexagons to
 ab-initio values\cite{georges3} and circles [a,b] to experimental values from Ref.\cite{stb,sperl}. (Note the circles with the cross denote the annealed
samples from Ref.\cite{sperl}).}
\label{fig.5}
\end{figure}
Now we proceed further and calculate the spin stiffness as a function of the concentration of Mn. As a first step the magnon energy $\omega(\bf {q})$ is plotted as a function
 of ${\bf q}^2$ for different $x$, as shown in fig.~\ref{fig.4}. Here $\omega(\rm{\bf q})$ is extracted from the first peak of $A({\bf q},\omega)$ in the (100) direction. The
system size varies from $L$=32 to $L$=44. The slope of these curves gives the spin stiffness $D(x)$ for
various $x$. Note that the perfect quadratic nature of the dispersion curves supports our claim that the average over disorder configurations is sufficient. In fig.~\ref{fig.5}
 we have plotted the spin stiffness $D(x)$ (in meV.{\AA}$^{2}$) as a function of Mn density. As stated before the lattice parameter used in our simple cubic system  
is $a$=$\frac{a_0}{4^{1/3}}$=3.55\AA, which is also used to calculate $D(x)$ here. It is worth noting that using the zinc blende lattice parameter (${a_0}$) here would lead to stiffness values $\sim$2.5 times
 larger than the present ones. Let us now discuss the results.  We have found that $D(x)$ is sensitive to the couplings
at large distances which appear to be strongly finite size dependent. Thus, the calculation of $D(x)$ is performed using the couplings from different system sizes ranging from
 $L$=16 to $L$=24, leading to error bars for $D(x)$ shown by the shaded region in the figure. It is to be noted that these couplings do not have a significant effect on the T$_C$s.
 In the figure we have shown the values of $D(x)$ obtained from ab-initio studies\cite{georges3}. Note that the stiffness values shown in Fig. 4 of Ref.\cite{georges3}
 actually correspond to $D$S where S=5/2\cite{correction}. Available experimental data for well annealed and as-grown samples\cite{stb,sperl} are also shown in the figure. Surprisingly
we find that the spin stiffness values from our model are in very good agreement, for the overall range of Mn concentration, with those obtained from first principle
studies. We also note that the agreement with the experimental results of Sperl \textit{et al.}\cite{sperl} is for the well annealed sample of thickness 200nm. This is
consistent with the fact that our couplings are calculated for optimally annealed samples. In Ref.~\cite{stb} the authors have reported a value of $D$$\approx$100 meV.\AA$^{2}$
 measured by ferromagnetic resonance in as-grown sample for $x$=0.05, which is also in good agreement with our results but then the error bar is about 40$\%$ and in all
 likelihood this stiffness should increase if the sample is optimally annealed. The deviation between our results and some experimental data could be explained
by the fact that our calculations are performed for optimally annealed samples.

Thus to conclude, we have provided a quantitative model study of the magnetic excitations in Mn doped GaAs. The model parameters were fixed from the first
principle calculations. We have calculated the magnon DOS, the``mobility edge'' separating the localized states from extended ones, the dynamical spectral function, the
spin stiffness and the T$_C$s as a function of the Mn concentration. We found a remarkable agreement between our theory and first principle studies for both
the spin stiffness as well as the  T$_C$s. At the same time we were able to reproduce most of the available experimental results. The strength of the model lies in the fact
that it can be applied to a whole family of DMS materials and this proves to be the link between model approaches and first principle calculations.

\acknowledgments
One of the authors, AC, wishes to thank the RTRA Nanosciences Fondation for financial support.

\end{document}